\documentclass[aps,prl,reprint,showpacs,floatfix]{revtex4-1}
\usepackage{amsmath}
\usepackage{slashed}
\usepackage{txfonts}
\usepackage{microtype}
\usepackage{graphicx}
\usepackage[breaklinks=true]{hyperref}
\usepackage{color}

\def\sout{\bgroup\markoverwith
{\textcolor{red}{\rule[0.5ex]{2pt}{0.5pt}}}\ULon}
\def\be{\begin{equation}}
\def\ee{\end{equation}}
\def\bes{\begin{equation*}}
\def\ees{\end{equation*}}
\def\bea{\begin{eqnarray}}
\def\eea{\end{eqnarray}}
\def\beas{\begin{eqnarray*}}
\def\eeas{\end{eqnarray*}}
\def\bal#1\eal{\begin{align}#1\end{align}}
\def\bals#1\eals{\begin{align*}#1\end{align*}}
\newcommand{\bra}[1]{\langle #1|}
\newcommand{\ket}[1]{|#1\rangle}
\newcommand{\braket}[2]{\langle #1|#2\rangle}
\newcommand{\bk}[1]{\langle #1\rangle}

\renewcommand{\vec}{\vectorsym}

\renewcommand*{\vec}[1]{\boldsymbol{#1}}

\graphicspath{{figures/}}

\usepackage[normalem]{ulem}

\begin{document}

\title{Anomalous screening of quantum impurities by a neutral environment}

\author{E. Yakaboylu}
\author{M. Lemeshko}

\affiliation{IST Austria (Institute of Science and Technology Austria), Am Campus 1, 3400 Klosterneuburg, Austria}

\date{\today}

\begin{abstract}

It is a common knowledge that an effective interaction of a quantum impurity with an electromagnetic field can be screened by surrounding charge carriers, whether mobile or static. Here we demonstrate that very strong, `anomalous' screening can take place in the presence of a neutral, weakly-polarizable environment, due to an exchange of orbital angular momentum between the impurity and the bath. Furthermore, we show that it is possible to generalize all phenomena related to isolated impurities in an external field  to the case when a many-body environment is present, by casting the problem in terms of the angulon quasiparticle. As a result, the relevant observables such as the effective Rabi frequency, geometric phase, and impurity spatial alignment are straightforward to evaluate in terms of a single parameter: the angular-momentum-dependent screening factor.

\end{abstract}

\maketitle

It is quite intuitive that once an impurity is immersed in a dielectric medium, its response to an external electromagnetic field is reduced -- or `screened' -- due to redistribution of charges in the dielectric~\cite{JacksonEM}. This classical description implies that if the medium is neutral and only weakly-polarizable, it induces a negligible change in impurity-light interactions, if any at all. Physics becomes more complicated, however, when quantum effects come into play. There, even the vacuum can behave as a medium with a finite dielectric permittivity due to virtual pair fluctuations, with non-linear effects taking place in the presence of strong electric fields~\cite{Piazza_2012}. These quantum fluctuations can also screen the impurity charge in a medium, as has been shown e.g. for graphene~\cite{Terekhov_2008}. Furthermore, due to the electron-phonon interactions, the Coulomb potential between two charged particles is screened in various settings, such as the jellium model~\cite{RevModPhys.65.677}. Another important example is the Kondo screening, where the dipole moments of magnetic impurities are screened by conduction electrons~\cite{kondo1964resistance, MedvedyevaPRB13}.

Here we uncover another type of screening -- that due to exchange of orbital angular momentum between the impurity and the surrounding quantum many-body bath. While such a screening takes place even for a bath `blind' to an electromagnetic field, it results in an anomalous decrease of the impurity susceptibility parameters, such as the effective dipole moment and polarizability. We start from the most general Hamiltonian describing an impurity interacting with a time-dependent electromagnetic (EM) field, which in the electric dipole approximation is given by:
\be
\label{free_hamiltonian}
\hat H_\text{imp-em}(t) = \hat  H_\text{imp} - \hat{\vec{d}} \cdot \vec{E}(t) \, .
\ee
Here $\hat H_\text{imp}$ is the Hamiltonian of the impurity, $\hat{\vec{d}}$ is its corresponding electric dipole operator, and $\vec{E}(t)$ is the electric field component of the EM field.  The simplest Hamiltonian for an impurity possessing orbital angular momentum is given by $\hat H_\text{imp} = B \hat{\vec{L}}^2$, where $\hat{\vec{L}}$ is the angular momentum operator. The constant $B$ depends on the particular system under investigation. For example, for the kinetic energy of a linear-rotor molecule, $B = 1/(2I)$ is the rotational constant with $I$ the moment of inertia~\cite{LevebvreBrionField2} (we use the units of $\hbar \equiv 1$ hereafter). For $t_{2g}$-electron orbitals in solids, $B = - \mathcal{J}/2$, where $\mathcal{J}$ parametrizes Hund's exchange coupling~\cite{georges2013}. Further degrees of freedom, such as electronic and nuclear spins, electron hopping, or a crystal field, will result in additional terms in $\hat H_\text{imp}$. For some other systems, such as highly-excited Rydberg electrons~\cite{GallagherRydbergAtoms}, or complex polyatomic molecules~\cite{BernathBook}, the impurity Hamiltonian might assume an overall different form. However, since the effects discussed in this paper originate from the orbital angular momentum transfer, the qualitative picture is not expected to change substantially.

In the presence of a neutral many-particle environment, the full Hamiltonian of the system is given by:
\be
\label{imp_hamiltonian}
\hat H(t) = \hat H_\text{imp-em}(t) + \hat H_\text{bath} + \hat H_\text{imp-bath} \, .
\ee
Note that we assume the environment to be weakly-polarizable, and therefore neglect its coupling to an external field. However, the impurity-bath interactions (of electrostatic, induction and dispersion type) are still present~\cite{StoneBook13}. We consider a neutral bosonic bath as described by the Hamiltonian, $\hat H_\text{bath} = \sum_{k \lambda \mu} \omega_k \hat{b}^\dagger_{k \lambda \mu} \hat{b}_{k \lambda \mu}$, with $\omega_k $ the dispersion relation. Here $\hat{b}^\dagger_{k \lambda \mu} $ and  $\hat{b}_{k \lambda \mu}$ are the bosonic creation and annihilation operators, $\sum_k \equiv \int d k$, and $k$, $\lambda$, and $\mu$ label the corresponding quantum numbers of linear momentum, angular momentum and its projection on the $z$-axis, respectively~\cite{Lemeshko_2015, Lemeshko_2016_book, PhysRevX.6.011012}. Such a bath can be represented e.g.\ by lattice phonons~\cite{Mahan90}, Bogoliubov excitations in a Bose-Einstein Condensate (BEC)~\cite{Pitaevskii2016}, or phonons, rotons, and ripplons in superfluid helium~\cite{StienkemeierJPB06}. For simplicity, in what follows we will refer to the bosonic excitations as `phonons.'   As it has been shown in Refs~\cite{Lemeshko_2015, Lemeshko_2016_book, PhysRevX.6.011012}, the interaction of an impurity carrying orbital angular momentum with a bosonic bath can be described as:
\be
\hat H_\text{imp-bath} =  \sum_{k \lambda \mu} U_\lambda (k) \left[ Y^{*}_{\lambda \mu}(\hat{\theta},\hat{\phi}) \hat{b}^\dagger_{k \lambda \mu} + Y_{\lambda \mu}(\hat{\theta},\hat{\phi}) \hat{b}_{k \lambda \mu}  \right] \, ,
\ee
where $U_\lambda (k)$ is the  angular-momentum-dependent coupling strength. As the interaction depends on the angle operators $\hat{\theta},\hat{\phi}$ of the impurity via the spherical harmonics $Y_{\lambda \mu}(\hat{\theta},\hat{\phi})$, the impurity in the angular state $\ket{jm}$ can undergo a transition to $\ket{j'm'}$ by absorption or emission of a phonon with the quantum numbers $k, \lambda, \mu$.  

\begin{figure*}
  \centering
  \includegraphics[width=\linewidth]{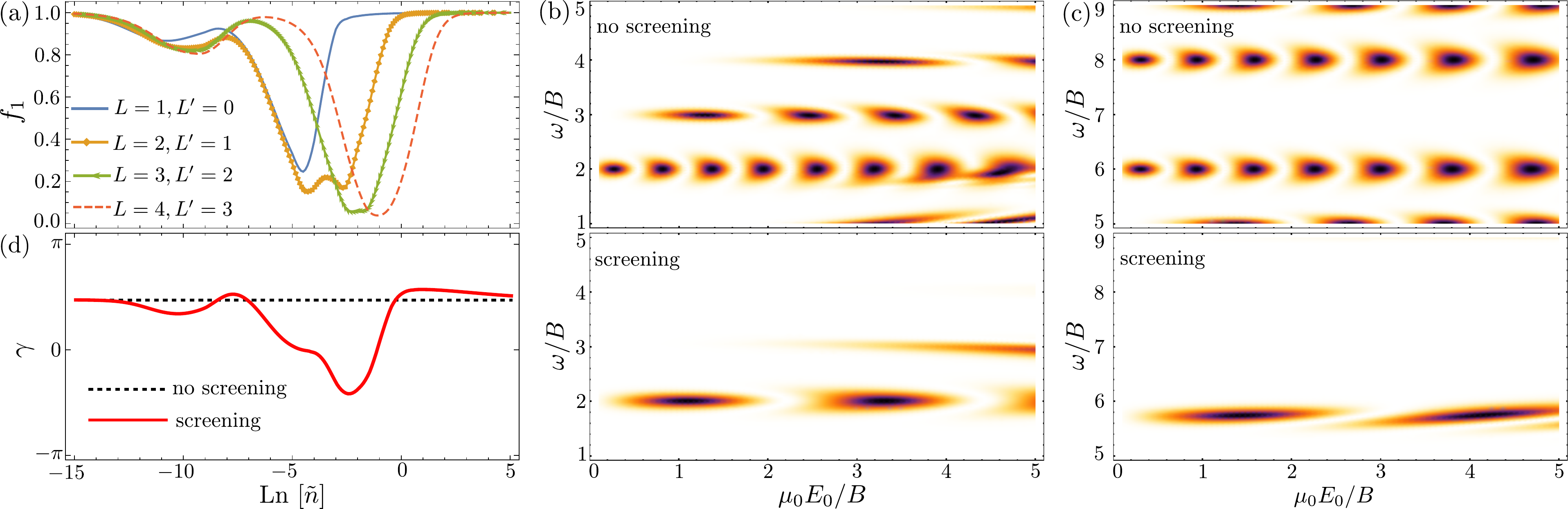}
 \caption{(Color online) (a)  The screening factor $f_1^{L,L'}$ for selected values of $L,L'$. (b) Total absorption of a free ground-state impurity (top), compared to a screened one, at  bath density $\text{Ln~}[\tilde{n}]=-4.5$ (bottom); (c) Total absorption of a free   impurity in the $L=3$ state (top), compared to a screened one, at   bath density $\text{Ln~}[\tilde{n}]=-1.0$ (bottom); (d) Geometric phase of the screened impurity (solid line) compared to that of a free impurity (dashed line), as  a function of bath density. See text.}
 \label{rabi}
\end{figure*}

In principle, it is extremely challenging to obtain exact time-dependent solutions to the full Hamiltonian of Eq.~(\ref{imp_hamiltonian}). The problem can be simplified tremendously, however, if one approaches it from the perspective of quasiparticles. Namely, it has been recently  shown that  impurities whose orbital angular momentum is coupled to a many-body bath form the angulon quasiparticles~\cite{Lemeshko_2015, Lemeshko_2016_book, PhysRevX.6.011012, Bikash16, Redchenko16, Li16}. This novel kind of quasiparticles can be thought of as a non-Abelian counterpart of polarons~\cite{Devreese15}, as it represents a quantum rotor dressed by a many-body bosonic field. Furthermore it was demonstrated that the predictions of the angulon theory are in good agreement with experiment for molecules in superfluid helium nanodroplets~\cite{lemeshko2016quasiparticle, Shepperson16}.

Accordingly, the full Hamiltonian of Eq.~\eqref{imp_hamiltonian} can be rewritten as $ \hat H(t) = \hat H_{A} - \hat{\vec{d}} \cdot \vec{E}(t) \otimes \mathbf{1} $, where $\hat H_A = \hat H_\text{imp} + \hat H_\text{bath} + \hat H_\text{imp-bath}$ is the angulon Hamiltonian, and the identity operator indicates that only the impurity interacts with the electric field. Taking only single-phonon excitations into account~\footnote{It is possible to construct a more general variational state taking higher order excitations into account~\cite{Levinsen_2015,Ngampruetikorn_2013,Parish_2013}. While this might result in even stronger anomalous screening, it is not expected to change the results qualitatively.}, the angulon eigenstate $\ket{A_{LM}}$ can be approximated by the following variational ansatz~\cite{Lemeshko_2015}:
\be
\label{angulon_states}
\ket{A_{LM}} = \sqrt{Z_{L}}\ket{0} \ket{ L M} + \sum_{k \lambda \mu j m} \beta^{L}_{k \lambda j} C^{L M}_{j m, \lambda \mu}  \hat{b}^\dagger_{k \lambda \mu} \ket{0} \ket{j m} \, ,
\ee
with $L$ and $M$ being the total angular momentum and its projection on the laboratory-frame $z$-axis, respectively. Here $\ket{0}$ represents the vacuum of bath excitations, $C^{L M}_{j m, \lambda \mu}$ are the Clebsch-Gordan coefficients~\cite{Varshalovich}, and $\sqrt{Z_{L}}$ and $\beta^{L}_{k \lambda j}$ are the variational parameters.  Eq.~\eqref{angulon_states} is straightforward to understand in the quasiparticle language: the first term corresponds to a bare impurity, with $Z_{L}$ being the quasiparticle weight, while the second term describes the field of many-particle excitations due to the impurity-bath interactions.

We start with the first-order expansion of the electric dipole operator, $ \hat{\vec{d}} \cdot \vec{E}(t) \approx \vec{\hat{\mu}_0} \cdot \vec{E}(t)$ (higher-order terms will be discussed below). Here $\vec{\hat{\mu}_0}$ is the permanent dipole moment operator of the impurity~\cite{Atkins2011molecular}. In the angulon basis, the state vector can be written as $\ket{\psi (t)} = \sum_{L M} K_{L M} (t) \ket{A_{L M}}$. The evolution of the corresponding amplitudes, $K_{LM}(t)$, is given by the Schr\"odinger equation:
\bal
\label{sch_eqn_angulon}
\nonumber i \frac{d K_{L M}}{d t} &  = - \sqrt{\frac{4\pi}{3}} \sum_{L' M' q}   K_{L' M'} \, E_q (t) |\vec{\mu_{0}}| \bra{L' M'}Y_{n q}(\hat{\theta},\hat{\phi})\ket{L M} f_n^{L,L'} \\
& + \varepsilon_L K_{L M} \, ,
\eal
where $\varepsilon_L = \bra{A_{LM}} \hat H_A \ket{A_{LM}} $, and $E_q$ with $q=\{0, \pm1 \}$ give the spherical components of $\vec{E}$. In Eq.~\eqref{sch_eqn_angulon} we separated out the factor,
\bal
 \nonumber f_n^{L,L'} & = \sqrt{Z_{L'}}^{*} \sqrt{Z_{L}} + \begin{pmatrix}
  L' & n & L \\
  0 & 0 & 0
\end{pmatrix}^{-1} \sum_{k \lambda j j'} \beta^{L' \, *}_{k \lambda j'} \beta^{L}_{k \lambda j} \begin{Bmatrix}
  j & \lambda & L \\
  L' & n & j'
\end{Bmatrix}\\
\label{f_funcs} & \times (-1)^{L+L' +\lambda+j'} \sqrt{2j'+1} C^{j\,0}_{j' \, 0 n 0}  \, ,
\eal
which we will refer to as the `angular-momentum-dependent screening factor.' The round and curly brackets in Eq.~\eqref{f_funcs} denote the Wigner $3j$-, and $6j$-symbols, respectively~\cite{Varshalovich}. We see that the same selection rules that applied to the angular momentum of the bare impurity,  now apply to the \textit{total} angular momentum of the angulon, $L$. Therefore,  Eq~(\ref{sch_eqn_angulon}) represents the Schr\"{o}dinger equation for a single particle -- the angulon -- interacting with an EM field. The only difference is that now the effective dipole moment, $ f_n^{L,L'} |\vec{\mu_{0}}|$, depends on the angular state of the impurity via the screening factor $ f_n^{L,L'}$, in analogy to the energy-dependent susceptibility of QED vacuum~\cite{peskin1995introduction}.

In the limit of $\beta^{L}_{k \lambda j} \to 0$, $Z_{L} \to 1$, and hence $f_n^{L,L'} \to 1$, Eq.~(\ref{sch_eqn_angulon}) reduces to the usual Schr\"odinger equation of an isolated impurity in an EM field. However, for non-vanishing $\beta_{k \lambda j}^L$, the screening factor $|f_n^{L,L'}| <1$~\footnote{See the Supplemental Material}: effective impurity-field interactions are proportional to the quasiparticle weight $Z_L$, which decreases if bath excitations are created.

In order to illustrate the effect of the bath on impurity-field interactions, we evaluate several observables, such as the effective Rabi frequency, geometric phase, and spatial alignment of the impurity axes.  Without loss of generality, we consider a bath with the Bogoliubov dispersion relation, $\omega_k = \sqrt{\epsilon_k (\epsilon_k + 2 g_{\text{bb}} n)}$~\cite{Pitaevskii2016}, where $\epsilon_k = k^2/(2m)$ with $m$ the boson mass and $n$ the boson particle density, and $g_{\text{bb}} = 4\pi a_{\text{bb}}/m$ where we set the boson-boson scattering length to $a_{\text{bb}} = 3.3 /\sqrt{m B}$. We choose the impurity-boson interaction as that derived for an ultracold molecule interacting with a dilute BEC, $U_\lambda (k) = \sqrt{8 n k^2 \epsilon_k/(\omega_k (2\lambda+1))} \int dr r^2 v_\lambda (r) j_\lambda (k r)$ where $j_\lambda (k r)$ is the spherical Bessel function~\cite{Lemeshko_2015}. We model the coupling using Gaussian functions, $v_\lambda (r) =  u_\lambda (2\pi)^{3/2} e^{-r^2/(2 r_\lambda^2)}$, and focus on the leading $\lambda$ orders, setting the parameters to $u_0 = 1.75 u_1 = 218 B$, and $r_0 = r_1 = 1.5 /\sqrt{m B}$. Taking into account higher-order couplings with $\lambda \ge 2$ will alter the selection rules on the boson-impurity scattering, however, is not expected to change the results qualitatively.

We study the behavior of the system as a function of  the dimensionless bath density, $\tilde{n} \equiv n (m B)^{-3/2}$, and for the sake of simplicity, we consider a linearly polarized EM field along the z-direction $E(t) = \mathcal{E}(t) \cos(\omega t) $ with the field frequency $\omega$, and the field envelope $\mathcal{E}(t)$. A linearly polarized field preserves cylindrical symmetry and renders $M$ a good quantum number. Here we focus on the $M=0$ manifold and omit the index $M$. We solve the Schr\"{o}dinger equation~(\ref{sch_eqn_angulon}) numerically taking into account terms up to $L_{\text{max}} = 50$ with the initial condition $K_{L} (t_i) = \delta_{L L_i} $.

In Fig.~\ref{rabi}(a) we present the screening factor for different angular-momentum states, as a function of the bath density. While for very low and very high densities the screening factor does not vary with $\tilde{n}$ and $L,L'$ substantially, there occur pronounced minima in the screening factor at intermediate densities. The latter correspond to the instabilities accompanied by the transfer of angular momentum from the impurity to the bath~\cite{Lemeshko_2015}. Such a drastic decrease in the screening factor is the manifestation of the anomalous screening.

%\begin{figure*}
%  \centering
%    \includegraphics[width=\linewidth]{fig2_h}
% \caption{(Color online) (a) The  screening factor $f_2^{L,L'}$, for selected values of $L,L'$. (b) Adiabatic alignment of a free $\mbox{CS}_2$ molecule   in a bath of selected densities, as illustrated by the time evolution of the alignment cosine; (c) Same as (b), for the case of  non-adiabatic alignment. See text.}
%  \label{allign}
%\end{figure*}

Let us now evaluate the total absorption of an impurity inside a neutral bath, as given by 
%$ \mathcal{T}_L = 1 - |\bra{A_L} U(t_f, t_i) \ket{A_L}|^2 $, where \rtext{$U(t_f, t_i) = ... $. \btext{[ML: please fill in!]} is the evolution operator}
$ \mathcal{T}_L = 1 - |\braket{A_L}{\psi(t_f)}|^2 $ with $\ket{\psi(t_i)} = \ket{A_L}$. 
Fig.~\ref{rabi}(b) shows $ \mathcal{T}_L$ as a function of the applied field energy, $\mu_0 E_0$, and the EM frequency, $\omega$, with and without a bath. The applied EM pulse is given by $\mathcal{E}(t) = E_0 \exp\left[-4 \ln(2) t^2/\tau^2\right]$, with the FWHM pulse duration $\tau = 6 \pi / B$ and the field amplitude  $E_0$. Close to the resonance, the dynamics is dominated by Rabi oscillations, which correspond to peaks in absorption (dark shade in the figure). Thus, the peaks at $\omega/B=2, 3,$ and $4$ correspond to the single-photon $L=0 \to L=1$ transition, two-photon  $L=0 \to L=2$ transition, and three-photon  $L=0 \to L=3$ transition, respectively. In the bottom panel of Fig.~\ref{rabi}(b), we see the result of anomalous screening -- a drastic decrease of the Rabi frequency. Accordingly, we can identify the effective Rabi frequency through  the  screening factor $f$: $\Omega_{L,L'}^A = f_1^{L,L'} \mu_0  E_0 \bra{L}\cos(\hat{\theta})\ket{L'} = f_1^{L,L'} \Omega_{L,L'} \, .$ For instance, for the $L=0 \to L=1$ transition the Rabi frequency is given by $ \Omega_{0,1}^A = \mu_0 E_0 f_1^{0,1} / \sqrt{3}$. At the instability density of $\text{Ln~}[\tilde{n}]=-4.5$, we obtain $f_1^{0,1} \approx 1/4 $, which is consistent with the plots shown in Fig~\ref{rabi}~(b). A similar behavior is observed for the total absorption for the impurity prepared in the third excited state, $L=3$, see Fig.~\ref{rabi}(c). We note that in the  regime of weak impurity-bath coupling, the energy splittings between the stable angulon states are close to the ones of an isolated impurity~[12]. As a result, the resonant frequencies for electromagnetic absorption are approximately the same.   

Another phenomenon we consider is the geometric phase accumulated during a cyclic evolution of the impurity~\cite{berry1984quantal,Aharonov_1987}. Following Aharanov and Anandan, any cyclic evolution may result in a geometric phase as given by $ \gamma = \phi + \int_0^\tau dt\, \bra{\psi (t)} \hat{\mathcal{H}}(t) \ket{\psi(t)} $, where the second term refers to the dynamical phase. Let us start from  one of the angulon eigenstates, $\ket{\psi_L (0)} = \ket{A_L}$, and let it evolve during a time interval $\tau$ into the same state up to a total phase, $ \ket{\psi_L (\tau)} = \exp(i \phi) \ket{A_L} $. The following parameters $\omega = 20 B$, $\mathcal{E}(t) = E_0 \sin^2 (\pi t/\tau)$, $\tau = 30/B$, and $E_0 = 11 B$ bring the system back to the initial state after the time $\tau$ for all densities. In Fig.~\ref{rabi} (d), we show the resulting geometric phase for the $L=1$ angulon state as a function of the bath density.

In order to get more insight into how a many-body environment influences the geometric phase, we consider a system of two levels, $L=0$ and $L=1$,  in a constant electric field. The corresponding Hamiltonian  can be written as $ \hat{\mathcal{H}} = \sigma_0 \, R_0 + \vec{\sigma} \cdot \vec{R} $, with some $R_0$ and $\vec{R}$, where  $\sigma_0$ and $\vec{\sigma}$ are the identity matrix and the vector of Pauli matrices, respectively. The time-evolution operator is given by:
\be
\hat U(t,0) = \exp(-i R_0 t)\left(\sigma_0 \cos(R t) - i \sin(R t) \, \vec{\sigma} \cdot \vec{R}/R \right) \, 
\ee
with $R \equiv |\vec{R}|$. The state evolution is cyclic under the period of $\tau = \pi /R$ up to the total phase $\phi = \pi (1 - R_0/R) $. The dynamical phase, on the other hand, is given by $- \int_0^\tau dt\,\bra{\psi_L (t)} \hat{\mathcal{H}} \ket{\psi_L (t)} = - \pi (R_0 \pm R_z)/R$, which leads to
%\be
%\beta = \pi \left( 1 + \frac{\varepsilon_0-\varepsilon_1}{\sqrt{(\varepsilon_0-\varepsilon_1)^2 + (2 f_1^{0,1} \mu_0 E_0  /\sqrt{3})^2}} \right) \, .
%\ee
\be
\gamma = \pi \left( 1 \pm (\varepsilon_0-\varepsilon_1)\left((\varepsilon_0-\varepsilon_1)^2 + (2 f_1^{0,1} \mu_0 E_0  /\sqrt{3})^2 \right)^{-1/2} \right) \, .
\ee
As for the Rabi frequency,  the neutral bath affects the geometric phase through the screening factor $f$. As a result, the geometric phase becomes density-dependent as shown in Fig.~\ref{rabi}(d). Note that $\gamma$ can assume both smaller and larger values compared to the isolated impurity case,  and vanishes identically for  certain densities.

\begin{figure}
  \centering
    \includegraphics[width=0.9\linewidth]{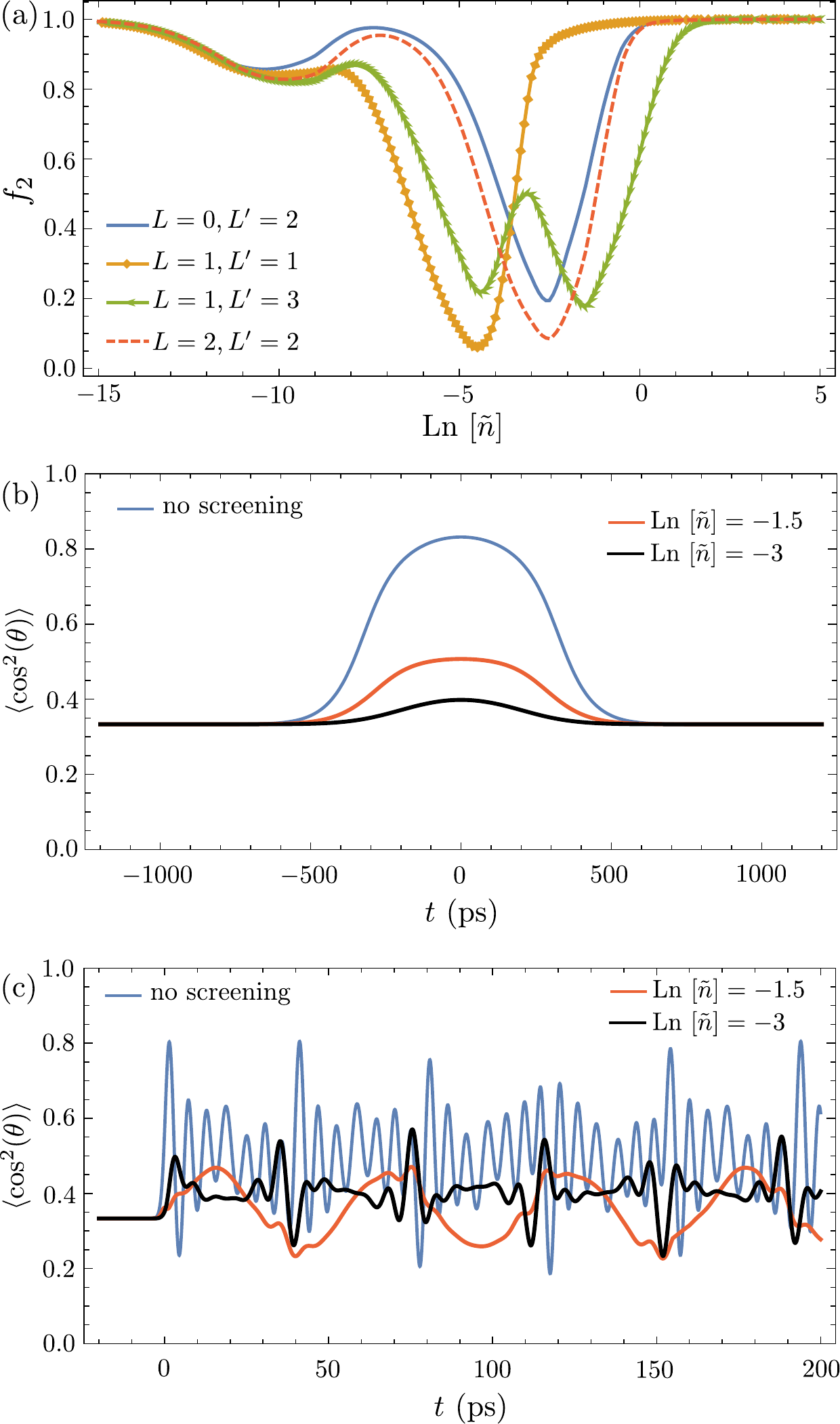}
 \caption{(Color online) (a) The  screening factor $f_2^{L,L'}$ for selected values of $L,L'$. (b) Adiabatic alignment of a free $\mbox{CS}_2$ molecule   in a bath of selected densities, as illustrated by the time evolution of the alignment cosine; (c) Same as (b), for the case of  non-adiabatic alignment. See text.}
  \label{allign}
\end{figure}

As a final example we consider effects of a neutral bath on the time-evolution of the impurity spatial alignment due to a  far-off-resonant laser pulse. Such a setting was realized e.g.\ in recent experiments on  adiabatic~\cite{PentlehnerPRA13}  and non-adiabatic~\cite{PentlehnerPRL13, ChristiansenPRA15, Shepperson16} molecular alignment in superfluid helium nanodroplets. Since in the case of intense off-resonant laser fields the second-order effects are important, we expand the dipole-field interaction as  $\hat{\vec{d}} \cdot \vec{E}(t) \approx \mu_0 E(t) \cos(\hat{\theta}) + (\Delta \alpha \cos^2(\hat{\theta}) + \alpha_\perp) E^2 (t) /2 $, where $\Delta \alpha = \alpha_\parallel -\alpha_\perp $ with $\alpha_\parallel$ and $\alpha_\perp$ being the polarizabilities in the direction parallel and perpendicular to the molecular axis. Furthermore,  far from any resonances,  the electric field can be averaged over the laser period so that the Hamiltonian is written in terms of the field envelope~\cite{friedrich1991spatial,friedrich1995polarization,Bretislav_1995}:
\be
\label{2nd_order_ham}
\hat H(t) = \hat H_A - \Delta \alpha \mathcal{E}^2(t) \tilde{Y}_{2,0}(\hat{\theta})/4 \otimes \mathbf{1}  \, ,
\ee
where $\tilde{Y}_{2,0}(\hat{\theta}) \equiv \sqrt{16 \pi/45} Y_{2,0}(\hat{\theta})$,  and the constant energy shifts are omitted. Similar to the permanent dipole case, the many-body Hamiltonian~(\ref{2nd_order_ham}) can be reduced to the single-particle Hamiltonian by introducing the screening factor, $f_2^{L,L'}$. The density dependence of the screening factor $f_2$ is shown in  Fig.~\ref{allign}(a).

Since an intense  laser field aligns the molecule along the direction of maximum polarizability~\cite{Stapelfeldt_2003,Torres_2005}, it is convenient to quantify the degree of alignment using the alignment cosine, $ \bk{\cos^2 (\hat{\theta}) } \equiv \bra{\psi(t)}\cos^2 (\hat{\theta})\otimes \mathbf{1} \ket{\psi(t)} $. If the pulse duration, $\tau$, is long compared to the rotational period, $T_\text{rot} =\pi/B$,  the alignment process is adiabatic. In such a case, the alignment cosine follows the electric field envelope. As an example, we consider a $\mbox{CS}_2$ molecule, whose parameters are given by: $\Delta \alpha = 67.5 \, \mbox{a.u.}$, $B = 4.97 \cdot 10^{-7} \, \mbox{a.u.} $ In Fig.~\ref{allign}(b) we compare the time evolution of $ \bk{\cos^2 (\hat{\theta}) }$ with the initial state $L=0$ for an adiabatic alignment of $\mbox{CS}_2$ inside an environment of various densities to that of an  isolated $\mbox{CS}_2$. We used the following parameters of the EM field: $\tau = 600 \, \mbox{ps}$,  the envelope $\mathcal{E}(t) = E_0 \exp\left[-4 \ln(2) t^2/\tau^2\right] $, and intensity $I = 1 \times 10^{10} \, \mbox{W}/\mbox{cm}^2$.  One can see that the screening manifests itself though a substantial reduction of the peak alignment. The magnitude of the screening  depends on the $f_2$-factor and can be derived analytically considering only two states, $L=0$ and $L=2$: 
%\oldtext{For adiabatic alignment the state vector can be defined as the instantaneous eigenstate of the Hamiltonian~(\ref{2nd_order_ham}), furthermore the peak value of the alignment corresponds to the peak of the EM field envelope. For example, for the $L=0$ state}
\be
\bk{\cos^2 (\hat{\theta}) }_\text{max} = \frac{1}{3} + \left(\frac{4 f_2^{2,2}}{21}\sin^2(\delta/2) + \frac{2 f_2^{0,2} }{3\sqrt{5}}\sin(\delta) \right) \, ,
\ee
where $\tan(\delta) = f_2^{0,2} \Delta \alpha E_0^2 / (\sqrt{45}(\varepsilon_2 - \varepsilon_0 - f_2^{2,2} \Delta \alpha E_0^2/21))$. As the screening factor $f_2^{0,2}$ decreases, the peak alignment decreases as well, as is the case for the density of $\text{Ln~}[\tilde{n}] = -3$, see Figs.~\ref{allign}(a) and (b).

%If the pulse duration $\tau$ is much smaller than the rotational period, the impurity-field interaction is non-adiabatic, which results in the revivals in the alignment cosine~\cite{OrtigosoFriedrich99, Lemeshko_2013_review}. Fig.~\ref{allign}(c) shows the time dependence of $\langle \cos^2(\hat{\theta}) \rangle$ following a short pulse with $\tau = 450 \, \mbox{fs}$ and $ I = 2.4 \times 10^{12} \, \mbox{W}/\mbox{cm}^2$ for a $\mbox{CS}_2$ molecule. While the frequency of the revivals in the presence of a bath  is similar to that of an isolated molecule, the maximum alignment scales with the screening factor $f_2$.

If $\tau \ll T_\text{rot}$, the impurity-field interaction is non-adiabatic, which results in the revivals in the alignment cosine~\cite{OrtigosoFriedrich99, Lemeshko_2013_review}. Note that in order for the pulse to be adiabatic with respect to the angulon formation, $\tau$ has to be long compared to the timescale of phonons in helium, $\tau_\text{ph}$. The latter is given by $\tau_\text{ph} = \mu^{-1}$, where   $\mu \sim k_B  \times 7.2$~K is the chemical potential of superfluid $^4$He~\cite{syvokon2006influence,ToenniesAngChem04}. This results in timescales $\tau_\text{ph} \sim 1$ ps for the typical response timescale of phonons. Accordingly, we use a pulse with $\tau = 4 \, \mbox{ps}$ and $ I = 1 \times 10^{11} \, \mbox{W}/\mbox{cm}^2$. Fig.~2 (c) shows the resulting time dependence of $\langle \cos^2(\hat{\theta}) \rangle$ for the case of $L=0$. While the frequency of the revivals in the presence of a bath  is similar to that of an isolated molecule, the maximum alignment scales with the screening factor $f_2$.

Thus, we have shown that a neutral weakly-polarizable  environment can induce a drastic screening of the impurity-field interactions due to the angular momentum transfer between the impurity and the bath. We developed a transparent analytic model based on the angulon quasiparticle, where all of the effects due to the bath are encapsulated in a single parameter -- the   screening factor $f$. Such a quasiparticle-based approach allows to extend the techniques developed for isolated atoms, molecules, and solid-state defects in external fields to the case when a many-particle environment is present.    The predicted effects should be measurable  with the state-of-the-art techniques used in quantum impurity experiments. For instance, the geometric phase can be measured using the impurity interference techniques~\cite{Hoffman,hanaguri2007quasiparticle,rutter2007scattering,jeon2014landau}, while experiments on molecules  rotating in superfluid helium nanodroplets allow to perform spectroscopic and alignment measurements~\cite{lehnig_rotational_2009, PentlehnerPRL13, PentlehnerPRA13, ChristiansenPRA15, Shepperson16, ToenniesAngChem04}. The presented formalism  can be generalized to the case of a fermionic environment, such as an ultracold degenerate Fermi gas~\cite{deMarcoSci99} or $^3$He~\cite{LeggettQuantLiquids,Pitaevskii2016}, as well as to Bose-Fermi mixtures~\cite{GunterPRL06}, which would further extend its domain of applicability.

\begin{acknowledgments}

We are grateful to G. Bighin, A. Deuchert, D. Forkert, S. Meuren, B. Midya, and R. Schmidt for valuable discussions. E. Y. acknowledges financial support received from the People Programme (Marie Curie Actions) of the European Union's Seventh Framework Programme (FP7/2007-2013) under REA grant agreement No. [291734]. This work was supported by the Austrian Science Fund (FWF), project Nr. P29902-N27.

\end{acknowledgments}

%\bibliography{angulon}

%

%\begin{widetext}
%\appendix
%\include{app}
%%\includepdf[pages=1-2]{app.pdf}
%\end{widetext}

\newpage

\begin{widetext}

\section{Supplemental Material \\ Anomalous screening of quantum impurities by a neutral environment}

Here we demonstrate that the screening factor cannot be larger than one. The screening factor is defined via
\be
\bra{A_{L' M'}}Y_{n q}(\hat{\theta},\hat{\phi}) \otimes \mathbf{1} \ket{A_{L M}}  = \bra{L' M'}Y_{n q}(\hat{\theta},\hat{\phi})\ket{L M} f_n^{L,L'} \, .
\ee
It follows from the Cauchy-Schwarz inequality that
\be
\label{eq:2}
| \bra{A_{L' M'}}Y_{n q}(\hat{\theta},\hat{\phi}) \otimes \mathbf{1} \ket{A_{L M}} |^2 \le \bra{A_{L M}}|Y_{n q}(\hat{\theta},\hat{\phi})|^2 \otimes \mathbf{1} \ket{A_{L M}} \ .
\ee
Using the variational state $\ket{A_{LM}}$, we can rewrite the right-hand side of Eq.~\eqref{eq:2} as:
\be
\label{eq3}
\bra{A_{L M}}|Y_{n q}(\hat{\theta},\hat{\phi})|^2 \otimes \mathbf{1} \ket{A_{L M}} = |Z_L| \bra{L M}|Y_{n q}(\hat{\theta},\hat{\phi})|^2 \ket{L M} + \sum_{k \lambda \mu j m j' m'} \beta_{k \lambda j'}^{L \, *} \beta_{k \lambda j}^{L} C^{LM}_{j' m' , \lambda \mu} C^{LM}_{j m , \lambda \mu} \bra{j'm'} |Y_{n q}(\hat{\theta},\hat{\phi})|^2 \ket{j m} \, .
\ee
With the help of the normalization condition for the variational state, $|Z_L| + \sum_{k \lambda j} |\beta_{k \lambda j}^L|^2 = 1$, Eq.~(\ref{eq3}) reads
\bal
\label{eq4}
\bra{A_{L M}}|Y_{n q}(\hat{\theta},\hat{\phi})|^2 \otimes \mathbf{1} \ket{A_{L M}} & =  \bra{L M}|Y_{n q}(\hat{\theta},\hat{\phi})|^2 \ket{L M} \\
\nonumber & + \sum_{k \lambda j j'} \beta_{k \lambda j'}^{L \, *} \beta_{k \lambda j}^{L} \left[ \bra{j'\lambda L M} |Y_{n q}(\hat{\theta},\hat{\phi})|^2 \otimes \mathbf{1} \ket{j \lambda L M} - \delta_{jj'} \bra{L M}|Y_{n q}(\hat{\theta},\hat{\phi})|^2 \ket{L M} \right] \, ,
\eal
with $\ket{j \lambda L M} \equiv \sum_{m,\mu} C^{LM}_{jm,\lambda\mu} \ket{jm ; \lambda \mu}$. From the Cauchy-Schwarz inequality it follows that

\bal
\sum_{k \lambda j j'} \beta_{k \lambda j'}^{L \, *} \beta_{k \lambda j}^{L}  \bra{j'\lambda L M} |Y_{n q}(\hat{\theta},\hat{\phi})|^2 \otimes \mathbf{1} \ket{j \lambda L M} & \le \sum_{k \lambda j} |\beta_{k \lambda j}^{L}|^2 \text{max} \left[ \bra{j'\lambda L M} |Y_{n q}(\hat{\theta},\hat{\phi})|^2 \otimes \mathbf{1} \ket{j \lambda L M} \right] \\
\nonumber &= \sum_{k \lambda j} |\beta_{k \lambda j}^{L}|^2 \bra{L M}|Y_{n q}(\hat{\theta},\hat{\phi})|^2 \ket{L M} \,,
\eal
where the second equality is due to the properties of the Clebsch-Gordan coefficients~\cite{Varshalovich1}. Thus, we obtain:
\be
\label{fund_ineq}
\bra{A_{L M}}|Y_{n q}(\hat{\theta},\hat{\phi})|^2 \otimes \mathbf{1} \ket{A_{L M}}  \le
\bra{L M}|Y_{n q}(\hat{\theta},\hat{\phi})|^2 \ket{L M} \, .
\ee

\vspace{0.5cm}
As a next step, we insert an identity, and rewrite the right-hand side of Eq.~\eqref{fund_ineq} as:
\be
\bra{L M} |Y_{n q}(\hat{\theta},\hat{\phi}) |^2 \ket{L M} = \sum_{L' M'} |\bra{L'M'} Y_{n q}(\hat{\theta},\hat{\phi})  \ket{L M}|^2 \,,
\ee
and similarly
\be
\bra{A_{LM}}|Y_{n q}(\hat{\theta},\hat{\phi})|^2 \otimes \mathbf{1} \ket{A_{L M}} |^2 = \sum_{L' M'} | \bra{A_{L' M'}}Y_{n q}(\hat{\theta},\hat{\phi}) \otimes \mathbf{1} \ket{A_{L M}} |^2 = \sum_{L' M'} |f_n^{L', L}|^2 |\bra{L'M'} Y_{n q}(\hat{\theta},\hat{\phi})  \ket{L M}|^2 \, .
\ee
From the inequality~(\ref{fund_ineq}), it follows that
\be
\sum_{L' M'} |\bra{L'M'} Y_{n q}(\hat{\theta},\hat{\phi})  \ket{L M}|^2 (|f_n^{L', L}|^2 -1) \le 0 \, ,
\ee
since $|\bra{L'M'} Y_{n q}(\hat{\theta},\hat{\phi})  \ket{L M}|^2$ are linearly independent, we provide $|f_n^{L', L}|^2 -1 \le 1$, and hence obtain that
\be
|f_n^{L', L}|  \le 1 \, .
\ee

It follows from Eq.~\ref{eq4} that the reason of the screening is the presence of the term $\beta$, i.e., the surrounding many-body environment. In the limit of $\beta \to 0$, we have $\bra{A_{L M}}|Y_{n q}(\hat{\theta},\hat{\phi})|^2 \otimes \mathbf{1} \ket{A_{L M}} =  \bra{L M}|Y_{n q}(\hat{\theta},\hat{\phi})|^2 \ket{L M}$. As $\beta$ increases, the inequality~\eqref{fund_ineq} increases as well. An increase of $\beta$, on the other hand, leads to a decrease in the quasi-particle weight $Z$ due to the normalization condition. Therefore, intuitively, due to the angular momentum transfer between the impurity and the surrounding bath,  the electric field interacts with `less of the impurity,' which results in a drop in the effective dipole moment.

%\bibliography{angulon}

\begin{thebibliography}{53}%
\makeatletter
\providecommand \@ifxundefined [1]{%
 \@ifx{#1\undefined}
}%
\providecommand \@ifnum [1]{%
 \ifnum #1\expandafter \@firstoftwo
 \else \expandafter \@secondoftwo
 \fi
}%
\providecommand \@ifx [1]{%
 \ifx #1\expandafter \@firstoftwo
 \else \expandafter \@secondoftwo
 \fi
}%
\providecommand \natexlab [1]{#1}%
\providecommand \enquote  [1]{``#1''}%
\providecommand \bibnamefont  [1]{#1}%
\providecommand \bibfnamefont [1]{#1}%
\providecommand \citenamefont [1]{#1}%
\providecommand \href@noop [0]{\@secondoftwo}%
\providecommand \href [0]{\begingroup \@sanitize@url \@href}%
\providecommand \@href[1]{\@@startlink{#1}\@@href}%
\providecommand \@@href[1]{\endgroup#1\@@endlink}%
\providecommand \@sanitize@url [0]{\catcode `\\12\catcode `\$12\catcode
  `\&12\catcode `\#12\catcode `\^12\catcode `\_12\catcode `\%12\relax}%
\providecommand \@@startlink[1]{}%
\providecommand \@@endlink[0]{}%
\providecommand \url  [0]{\begingroup\@sanitize@url \@url }%
\providecommand \@url [1]{\endgroup\@href {#1}{\urlprefix }}%
\providecommand \urlprefix  [0]{URL }%
\providecommand \Eprint [0]{\href }%
\providecommand \doibase [0]{http://dx.doi.org/}%
\providecommand \selectlanguage [0]{\@gobble}%
\providecommand \bibinfo  [0]{\@secondoftwo}%
\providecommand \bibfield  [0]{\@secondoftwo}%
\providecommand \translation [1]{[#1]}%
\providecommand \BibitemOpen [0]{}%
\providecommand \bibitemStop [0]{}%
\providecommand \bibitemNoStop [0]{.\EOS\space}%
\providecommand \EOS [0]{\spacefactor3000\relax}%
\providecommand \BibitemShut  [1]{\csname bibitem#1\endcsname}%
\let\auto@bib@innerbib\@empty
%</preamble>
\bibitem [{\citenamefont {Jackson}(1962)}]{JacksonEM}%
  \BibitemOpen
  \bibfield  {author} {\bibinfo {author} {\bibfnamefont {J.~D.}\ \bibnamefont
  {Jackson}},\ }\href@noop {} {\emph {\bibinfo {title} {Classical
  Electrodynamics}}}\ (\bibinfo  {publisher} {John Wiley and Sons},\ \bibinfo
  {year} {1962})\BibitemShut {NoStop}%
\bibitem [{\citenamefont {Di~Piazza}\ \emph {et~al.}(2012)\citenamefont
  {Di~Piazza}, \citenamefont {M\"uller}, \citenamefont {Hatsagortsyan},\ and\
  \citenamefont {Keitel}}]{Piazza_2012}%
  \BibitemOpen
  \bibfield  {author} {\bibinfo {author} {\bibfnamefont {A.}~\bibnamefont
  {Di~Piazza}}, \bibinfo {author} {\bibfnamefont {C.}~\bibnamefont {M\"uller}},
  \bibinfo {author} {\bibfnamefont {K.~Z.}\ \bibnamefont {Hatsagortsyan}}, \
  and\ \bibinfo {author} {\bibfnamefont {C.~H.}\ \bibnamefont {Keitel}},\
  }\href {\doibase 10.1103/RevModPhys.84.1177} {\bibfield  {journal} {\bibinfo
  {journal} {Rev. Mod. Phys.}\ }\textbf {\bibinfo {volume} {84}},\ \bibinfo
  {pages} {1177} (\bibinfo {year} {2012})}\BibitemShut {NoStop}%
\bibitem [{\citenamefont {Terekhov}\ \emph {et~al.}(2008)\citenamefont
  {Terekhov}, \citenamefont {Milstein}, \citenamefont {Kotov},\ and\
  \citenamefont {Sushkov}}]{Terekhov_2008}%
  \BibitemOpen
  \bibfield  {author} {\bibinfo {author} {\bibfnamefont {I.~S.}\ \bibnamefont
  {Terekhov}}, \bibinfo {author} {\bibfnamefont {A.~I.}\ \bibnamefont
  {Milstein}}, \bibinfo {author} {\bibfnamefont {V.~N.}\ \bibnamefont {Kotov}},
  \ and\ \bibinfo {author} {\bibfnamefont {O.~P.}\ \bibnamefont {Sushkov}},\
  }\href {\doibase 10.1103/PhysRevLett.100.076803} {\bibfield  {journal}
  {\bibinfo  {journal} {Phys. Rev. Lett.}\ }\textbf {\bibinfo {volume} {100}},\
  \bibinfo {pages} {076803} (\bibinfo {year} {2008})}\BibitemShut {NoStop}%
\bibitem [{\citenamefont {Brack}(1993)}]{RevModPhys.65.677}%
  \BibitemOpen
  \bibfield  {author} {\bibinfo {author} {\bibfnamefont {M.}~\bibnamefont
  {Brack}},\ }\href {\doibase 10.1103/RevModPhys.65.677} {\bibfield  {journal}
  {\bibinfo  {journal} {Rev. Mod. Phys.}\ }\textbf {\bibinfo {volume} {65}},\
  \bibinfo {pages} {677} (\bibinfo {year} {1993})}\BibitemShut {NoStop}%
\bibitem [{\citenamefont {Kondo}(1964)}]{kondo1964resistance}%
  \BibitemOpen
  \bibfield  {author} {\bibinfo {author} {\bibfnamefont {J.}~\bibnamefont
  {Kondo}},\ }\href@noop {} {\bibfield  {journal} {\bibinfo  {journal}
  {Progress of theoretical physics}\ }\textbf {\bibinfo {volume} {32}},\
  \bibinfo {pages} {37} (\bibinfo {year} {1964})}\BibitemShut {NoStop}%
\bibitem [{\citenamefont {Medvedyeva}\ \emph {et~al.}(2013)\citenamefont
  {Medvedyeva}, \citenamefont {Hoffmann},\ and\ \citenamefont
  {Kehrein}}]{MedvedyevaPRB13}%
  \BibitemOpen
  \bibfield  {author} {\bibinfo {author} {\bibfnamefont {M.}~\bibnamefont
  {Medvedyeva}}, \bibinfo {author} {\bibfnamefont {A.}~\bibnamefont
  {Hoffmann}}, \ and\ \bibinfo {author} {\bibfnamefont {S.}~\bibnamefont
  {Kehrein}},\ }\href@noop {} {\bibfield  {journal} {\bibinfo  {journal} {Phys.
  Rev. B}\ }\textbf {\bibinfo {volume} {88}},\ \bibinfo {pages} {094306}
  (\bibinfo {year} {2013})}\BibitemShut {NoStop}%
\bibitem [{\citenamefont {Lefebvre-Brion}\ and\ \citenamefont
  {Field}(2004)}]{LevebvreBrionField2}%
  \BibitemOpen
  \bibfield  {author} {\bibinfo {author} {\bibfnamefont {H.}~\bibnamefont
  {Lefebvre-Brion}}\ and\ \bibinfo {author} {\bibfnamefont {R.~W.}\
  \bibnamefont {Field}},\ }\href@noop {} {\emph {\bibinfo {title} {The Spectra
  and Dynamics of Diatomic Molecules}}}\ (\bibinfo  {publisher} {Elsevier, New
  York},\ \bibinfo {year} {2004})\BibitemShut {NoStop}%
\bibitem [{\citenamefont {Georges}\ \emph {et~al.}(2013)\citenamefont
  {Georges}, \citenamefont {de' Medici},\ and\ \citenamefont
  {Mravlje}}]{georges2013}%
  \BibitemOpen
  \bibfield  {author} {\bibinfo {author} {\bibfnamefont {A.}~\bibnamefont
  {Georges}}, \bibinfo {author} {\bibfnamefont {L.}~\bibnamefont {de' Medici}},
  \ and\ \bibinfo {author} {\bibfnamefont {J.}~\bibnamefont {Mravlje}},\
  }\href@noop {} {\bibfield  {journal} {\bibinfo  {journal} {Annu. Rev.
  Condens. Matter Phys.}\ }\textbf {\bibinfo {volume} {4}},\ \bibinfo {pages}
  {137} (\bibinfo {year} {2013})}\BibitemShut {NoStop}%
\bibitem [{\citenamefont {Gallagher}(2005)}]{GallagherRydbergAtoms}%
  \BibitemOpen
  \bibfield  {author} {\bibinfo {author} {\bibfnamefont {T.~F.}\ \bibnamefont
  {Gallagher}},\ }\href@noop {} {\emph {\bibinfo {title} {Rydberg atoms}}}\
  (\bibinfo  {publisher} {Cambridge University Press},\ \bibinfo {year}
  {2005})\BibitemShut {NoStop}%
\bibitem [{\citenamefont {Bernath}(2005)}]{BernathBook}%
  \BibitemOpen
  \bibfield  {author} {\bibinfo {author} {\bibfnamefont {P.~F.}\ \bibnamefont
  {Bernath}},\ }\href@noop {} {\emph {\bibinfo {title} {Spectra of atoms and
  molecules}}},\ \bibinfo {edition} {2nd}\ ed.\ (\bibinfo  {publisher} {Oxford
  University Press},\ \bibinfo {year} {2005})\BibitemShut {NoStop}%
\bibitem [{\citenamefont {Stone}(2013)}]{StoneBook13}%
  \BibitemOpen
  \bibfield  {author} {\bibinfo {author} {\bibfnamefont {A.}~\bibnamefont
  {Stone}},\ }\href@noop {} {\emph {\bibinfo {title} {The Theory of
  Intermolecular Forces}}}\ (\bibinfo  {publisher} {Oxford University Press},\
  \bibinfo {year} {2013})\BibitemShut {NoStop}%
\bibitem [{\citenamefont {Schmidt}\ and\ \citenamefont
  {Lemeshko}(2015)}]{Lemeshko_2015}%
  \BibitemOpen
  \bibfield  {author} {\bibinfo {author} {\bibfnamefont {R.}~\bibnamefont
  {Schmidt}}\ and\ \bibinfo {author} {\bibfnamefont {M.}~\bibnamefont
  {Lemeshko}},\ }\href {\doibase 10.1103/PhysRevLett.114.203001} {\bibfield
  {journal} {\bibinfo  {journal} {Phys. Rev. Lett.}\ }\textbf {\bibinfo
  {volume} {114}},\ \bibinfo {pages} {203001} (\bibinfo {year}
  {2015})}\BibitemShut {NoStop}%
\bibitem [{\citenamefont {Lemeshko}\ and\ \citenamefont
  {Schmidt}(2017)}]{Lemeshko_2016_book}%
  \BibitemOpen
  \bibfield  {author} {\bibinfo {author} {\bibfnamefont {M.}~\bibnamefont
  {Lemeshko}}\ and\ \bibinfo {author} {\bibfnamefont {R.}~\bibnamefont
  {Schmidt}},\ }\href@noop {} {\emph {\bibinfo {title} {Molecular impurities
  interacting with a many-particle environment: from ultracold gases to helium
  nanodroplets, book chapter in "Low Energy and Low Temperature Molecular
  Scattering" edited by A. Osterwalder and O. Dulieu}}}\ (\bibinfo  {publisher}
  {Royal Society of Chemistry},\ \bibinfo {year} {2017})\BibitemShut {NoStop}%
\bibitem [{\citenamefont {Schmidt}\ and\ \citenamefont
  {Lemeshko}(2016)}]{PhysRevX.6.011012}%
  \BibitemOpen
  \bibfield  {author} {\bibinfo {author} {\bibfnamefont {R.}~\bibnamefont
  {Schmidt}}\ and\ \bibinfo {author} {\bibfnamefont {M.}~\bibnamefont
  {Lemeshko}},\ }\href {\doibase 10.1103/PhysRevX.6.011012} {\bibfield
  {journal} {\bibinfo  {journal} {Phys. Rev. X}\ }\textbf {\bibinfo {volume}
  {6}},\ \bibinfo {pages} {011012} (\bibinfo {year} {2016})}\BibitemShut
  {NoStop}%
\bibitem [{\citenamefont {Mahan}(1990)}]{Mahan90}%
  \BibitemOpen
  \bibfield  {author} {\bibinfo {author} {\bibfnamefont {G.~D.}\ \bibnamefont
  {Mahan}},\ }\href@noop {} {\emph {\bibinfo {title} {Many-particle
  physics}}},\ Physics of solids and liquids\ (\bibinfo  {publisher} {Plenum},\
  \bibinfo {address} {New York, NY},\ \bibinfo {year} {1990})\BibitemShut
  {NoStop}%
\bibitem [{\citenamefont {Pitaevskii}\ and\ \citenamefont
  {Stringari}(2016)}]{Pitaevskii2016}%
  \BibitemOpen
  \bibfield  {author} {\bibinfo {author} {\bibfnamefont {L.~P.}\ \bibnamefont
  {Pitaevskii}}\ and\ \bibinfo {author} {\bibfnamefont {S.}~\bibnamefont
  {Stringari}},\ }\href@noop {} {\emph {\bibinfo {title} {Bose-Einstein
  Condensation and Superfluidity}}}\ (\bibinfo  {publisher} {Oxford University
  Press},\ \bibinfo {year} {2016})\BibitemShut {NoStop}%
\bibitem [{\citenamefont {Stienkemeier}\ and\ \citenamefont
  {Lehmann}(2006)}]{StienkemeierJPB06}%
  \BibitemOpen
  \bibfield  {author} {\bibinfo {author} {\bibfnamefont {F.}~\bibnamefont
  {Stienkemeier}}\ and\ \bibinfo {author} {\bibfnamefont {K.~K.}\ \bibnamefont
  {Lehmann}},\ }\href@noop {} {\bibfield  {journal} {\bibinfo  {journal} {J.
  Phys. B}\ }\textbf {\bibinfo {volume} {39}},\ \bibinfo {pages} {R127}
  (\bibinfo {year} {2006})}\BibitemShut {NoStop}%
\bibitem [{\citenamefont {Midya}\ \emph {et~al.}(2016)\citenamefont {Midya},
  \citenamefont {Tomza}, \citenamefont {Schmidt},\ and\ \citenamefont
  {Lemeshko}}]{Bikash16}%
  \BibitemOpen
  \bibfield  {author} {\bibinfo {author} {\bibfnamefont {B.}~\bibnamefont
  {Midya}}, \bibinfo {author} {\bibfnamefont {M.}~\bibnamefont {Tomza}},
  \bibinfo {author} {\bibfnamefont {R.}~\bibnamefont {Schmidt}}, \ and\
  \bibinfo {author} {\bibfnamefont {M.}~\bibnamefont {Lemeshko}},\ }\href@noop
  {} {\bibfield  {journal} {\bibinfo  {journal} {Phys. Rev. A}\ }\textbf
  {\bibinfo {volume} {94}},\ \bibinfo {pages} {041601(R)} (\bibinfo {year}
  {2016})}\BibitemShut {NoStop}%
\bibitem [{\citenamefont {Redchenko}\ and\ \citenamefont
  {Lemeshko}(2016)}]{Redchenko16}%
  \BibitemOpen
  \bibfield  {author} {\bibinfo {author} {\bibfnamefont {E.~S.}\ \bibnamefont
  {Redchenko}}\ and\ \bibinfo {author} {\bibfnamefont {M.}~\bibnamefont
  {Lemeshko}},\ }\href@noop {} {\bibfield  {journal} {\bibinfo  {journal}
  {Chem. Phys. Chem.}\ }\textbf {\bibinfo {volume} {17}},\ \bibinfo {pages}
  {3649} (\bibinfo {year} {2016})}\BibitemShut {NoStop}%
\bibitem [{\citenamefont {Li}\ \emph {et~al.}(2016)\citenamefont {Li},
  \citenamefont {Seiringer},\ and\ \citenamefont {Lemeshko}}]{Li16}%
  \BibitemOpen
  \bibfield  {author} {\bibinfo {author} {\bibfnamefont {X.}~\bibnamefont
  {Li}}, \bibinfo {author} {\bibfnamefont {R.}~\bibnamefont {Seiringer}}, \
  and\ \bibinfo {author} {\bibfnamefont {M.}~\bibnamefont {Lemeshko}},\
  }\href@noop {} {\bibfield  {journal} {\bibinfo  {journal} {arXiv:
  1610.04908}\ } (\bibinfo {year} {2016})}\BibitemShut {NoStop}%
\bibitem [{\citenamefont {Devreese}(2015)}]{Devreese15}%
  \BibitemOpen
  \bibfield  {author} {\bibinfo {author} {\bibfnamefont {J.~T.}\ \bibnamefont
  {Devreese}},\ }\href@noop {} {\bibfield  {journal} {\bibinfo  {journal}
  {arXiv:1012.4576}\ } (\bibinfo {year} {2015})}\BibitemShut {NoStop}%
\bibitem [{\citenamefont {Lemeshko}(2017)}]{lemeshko2016quasiparticle}%
  \BibitemOpen
  \bibfield  {author} {\bibinfo {author} {\bibfnamefont {M.}~\bibnamefont
  {Lemeshko}},\ }\href@noop {} {\bibfield  {journal} {\bibinfo  {journal}
  {Phys. Rev. Lett., in press; arXiv:1610.01604}\ } (\bibinfo {year}
  {2017})}\BibitemShut {NoStop}%
\bibitem [{\citenamefont {Shepperson}\ \emph {et~al.}(2017)\citenamefont
  {Shepperson}, \citenamefont {S{\o}ndergaard}, \citenamefont {Christiansen},
  \citenamefont {Kaczmarczyk}, \citenamefont {Zillich}, \citenamefont
  {Lemeshko},\ and\ \citenamefont {Stapelfeldt}}]{Shepperson16}%
  \BibitemOpen
  \bibfield  {author} {\bibinfo {author} {\bibfnamefont {B.}~\bibnamefont
  {Shepperson}}, \bibinfo {author} {\bibfnamefont {A.~A.}\ \bibnamefont
  {S{\o}ndergaard}}, \bibinfo {author} {\bibfnamefont {L.}~\bibnamefont
  {Christiansen}}, \bibinfo {author} {\bibfnamefont {J.}~\bibnamefont
  {Kaczmarczyk}}, \bibinfo {author} {\bibfnamefont {R.~E.}\ \bibnamefont
  {Zillich}}, \bibinfo {author} {\bibfnamefont {M.}~\bibnamefont {Lemeshko}}, \
  and\ \bibinfo {author} {\bibfnamefont {H.}~\bibnamefont {Stapelfeldt}},\
  }\href@noop {} {\bibfield  {journal} {\bibinfo  {journal} {arXiv:1702.01977}\
  } (\bibinfo {year} {2017})}\BibitemShut {NoStop}%
\bibitem [{Note1()}]{Note1}%
  \BibitemOpen
  \bibinfo {note} {It is possible to construct a more general variational state
  taking higher order excitations into account~\cite
  {Levinsen_2015,Ngampruetikorn_2013,Parish_2013}. While this might result in
  even stronger anomalous screening, it is not expected to change the results
  qualitatively.}\BibitemShut {Stop}%
\bibitem [{\citenamefont {Varshalovich}\ \emph {et~al.}(1988)\citenamefont
  {Varshalovich}, \citenamefont {Moskalev},\ and\ \citenamefont
  {Khersonskii}}]{Varshalovich}%
  \BibitemOpen
  \bibfield  {author} {\bibinfo {author} {\bibfnamefont {D.~A.}\ \bibnamefont
  {Varshalovich}}, \bibinfo {author} {\bibfnamefont {A.}~\bibnamefont
  {Moskalev}}, \ and\ \bibinfo {author} {\bibfnamefont {V.}~\bibnamefont
  {Khersonskii}},\ }\href@noop {} {\emph {\bibinfo {title} {Quantum theory of
  angular momentum}}}\ (\bibinfo  {publisher} {World Scientific},\ \bibinfo
  {year} {1988})\BibitemShut {NoStop}%
\bibitem [{\citenamefont {Atkins}\ and\ \citenamefont
  {Friedman}(2011)}]{Atkins2011molecular}%
  \BibitemOpen
  \bibfield  {author} {\bibinfo {author} {\bibfnamefont {P.~W.}\ \bibnamefont
  {Atkins}}\ and\ \bibinfo {author} {\bibfnamefont {R.~S.}\ \bibnamefont
  {Friedman}},\ }\href@noop {} {\emph {\bibinfo {title} {Molecular quantum
  mechanics}}}\ (\bibinfo  {publisher} {Oxford university press},\ \bibinfo
  {year} {2011})\BibitemShut {NoStop}%
\bibitem [{\citenamefont {Peskin}\ and\ \citenamefont
  {Schroeder}(1995)}]{peskin1995introduction}%
  \BibitemOpen
  \bibfield  {author} {\bibinfo {author} {\bibfnamefont {M.}~\bibnamefont
  {Peskin}}\ and\ \bibinfo {author} {\bibfnamefont {D.}~\bibnamefont
  {Schroeder}},\ }\href@noop {} {\emph {\bibinfo {title} {An introduction to
  quantum field theory}}}\ (\bibinfo  {publisher} {Westview Press},\ \bibinfo
  {year} {1995})\BibitemShut {NoStop}%
\bibitem [{Note2()}]{Note2}%
  \BibitemOpen
  \bibinfo {note} {See the Supplemental Material}\BibitemShut {NoStop}%
\bibitem [{\citenamefont {Berry}(1984)}]{berry1984quantal}%
  \BibitemOpen
  \bibfield  {author} {\bibinfo {author} {\bibfnamefont {M.~V.}\ \bibnamefont
  {Berry}},\ }\href@noop {} {\bibfield  {journal} {\bibinfo  {journal}
  {Proceedings of the Royal Society of London. A. Mathematical and Physical
  Sciences}\ }\textbf {\bibinfo {volume} {392}},\ \bibinfo {pages} {45}
  (\bibinfo {year} {1984})}\BibitemShut {NoStop}%
\bibitem [{\citenamefont {Aharonov}\ and\ \citenamefont
  {Anandan}(1987)}]{Aharonov_1987}%
  \BibitemOpen
  \bibfield  {author} {\bibinfo {author} {\bibfnamefont {Y.}~\bibnamefont
  {Aharonov}}\ and\ \bibinfo {author} {\bibfnamefont {J.}~\bibnamefont
  {Anandan}},\ }\href {\doibase 10.1103/PhysRevLett.58.1593} {\bibfield
  {journal} {\bibinfo  {journal} {Phys. Rev. Lett.}\ }\textbf {\bibinfo
  {volume} {58}},\ \bibinfo {pages} {1593} (\bibinfo {year}
  {1987})}\BibitemShut {NoStop}%
\bibitem [{\citenamefont {Pentlehner}\ \emph
  {et~al.}(2013{\natexlab{a}})\citenamefont {Pentlehner}, \citenamefont
  {Nielsen}, \citenamefont {Christiansen}, \citenamefont {Slenczka},\ and\
  \citenamefont {Stapelfeldt}}]{PentlehnerPRA13}%
  \BibitemOpen
  \bibfield  {author} {\bibinfo {author} {\bibfnamefont {D.}~\bibnamefont
  {Pentlehner}}, \bibinfo {author} {\bibfnamefont {J.~H.}\ \bibnamefont
  {Nielsen}}, \bibinfo {author} {\bibfnamefont {L.}~\bibnamefont
  {Christiansen}}, \bibinfo {author} {\bibfnamefont {A.}~\bibnamefont
  {Slenczka}}, \ and\ \bibinfo {author} {\bibfnamefont {H.}~\bibnamefont
  {Stapelfeldt}},\ }\href@noop {} {\bibfield  {journal} {\bibinfo  {journal}
  {Physical Review A}\ }\textbf {\bibinfo {volume} {87}},\ \bibinfo {pages}
  {063401} (\bibinfo {year} {2013}{\natexlab{a}})}\BibitemShut {NoStop}%
\bibitem [{\citenamefont {Pentlehner}\ \emph
  {et~al.}(2013{\natexlab{b}})\citenamefont {Pentlehner}, \citenamefont
  {Nielsen}, \citenamefont {Slenczka}, \citenamefont {M{\o}lmer},\ and\
  \citenamefont {Stapelfeldt}}]{PentlehnerPRL13}%
  \BibitemOpen
  \bibfield  {author} {\bibinfo {author} {\bibfnamefont {D.}~\bibnamefont
  {Pentlehner}}, \bibinfo {author} {\bibfnamefont {J.~H.}\ \bibnamefont
  {Nielsen}}, \bibinfo {author} {\bibfnamefont {A.}~\bibnamefont {Slenczka}},
  \bibinfo {author} {\bibfnamefont {K.}~\bibnamefont {M{\o}lmer}}, \ and\
  \bibinfo {author} {\bibfnamefont {H.}~\bibnamefont {Stapelfeldt}},\
  }\href@noop {} {\bibfield  {journal} {\bibinfo  {journal} {Phys. Rev. Lett.}\
  }\textbf {\bibinfo {volume} {110}},\ \bibinfo {pages} {093002} (\bibinfo
  {year} {2013}{\natexlab{b}})}\BibitemShut {NoStop}%
\bibitem [{\citenamefont {Christiansen}\ \emph {et~al.}(2015)\citenamefont
  {Christiansen}, \citenamefont {Nielsen}, \citenamefont {Tobias},
  \citenamefont {Pentlehner}, \citenamefont {Underwood},\ and\ \citenamefont
  {Stapelfeldt}}]{ChristiansenPRA15}%
  \BibitemOpen
  \bibfield  {author} {\bibinfo {author} {\bibfnamefont {L.}~\bibnamefont
  {Christiansen}}, \bibinfo {author} {\bibfnamefont {J.~H.}\ \bibnamefont
  {Nielsen}}, \bibinfo {author} {\bibfnamefont {D.}~\bibnamefont {Tobias}},
  \bibinfo {author} {\bibfnamefont {V.}~\bibnamefont {Pentlehner}}, \bibinfo
  {author} {\bibfnamefont {J.~G.}\ \bibnamefont {Underwood}}, \ and\ \bibinfo
  {author} {\bibfnamefont {H.}~\bibnamefont {Stapelfeldt}},\ }\href@noop {}
  {\bibfield  {journal} {\bibinfo  {journal} {Phys. Rev. A}\ }\textbf {\bibinfo
  {volume} {92}},\ \bibinfo {pages} {1050} (\bibinfo {year}
  {2015})}\BibitemShut {NoStop}%
\bibitem [{\citenamefont {Friedrich}\ and\ \citenamefont
  {Herschbach}(1991)}]{friedrich1991spatial}%
  \BibitemOpen
  \bibfield  {author} {\bibinfo {author} {\bibfnamefont {B.}~\bibnamefont
  {Friedrich}}\ and\ \bibinfo {author} {\bibfnamefont {D.~R.}\ \bibnamefont
  {Herschbach}},\ }\href@noop {} {\bibfield  {journal} {\bibinfo  {journal}
  {Nature}\ }\textbf {\bibinfo {volume} {353}},\ \bibinfo {pages} {412}
  (\bibinfo {year} {1991})}\BibitemShut {NoStop}%
\bibitem [{\citenamefont {Friedrich}\ and\ \citenamefont
  {Herschbach}(1995{\natexlab{a}})}]{friedrich1995polarization}%
  \BibitemOpen
  \bibfield  {author} {\bibinfo {author} {\bibfnamefont {B.}~\bibnamefont
  {Friedrich}}\ and\ \bibinfo {author} {\bibfnamefont {D.}~\bibnamefont
  {Herschbach}},\ }\href@noop {} {\bibfield  {journal} {\bibinfo  {journal}
  {The Journal of Physical Chemistry}\ }\textbf {\bibinfo {volume} {99}},\
  \bibinfo {pages} {15686} (\bibinfo {year} {1995}{\natexlab{a}})}\BibitemShut
  {NoStop}%
\bibitem [{\citenamefont {Friedrich}\ and\ \citenamefont
  {Herschbach}(1995{\natexlab{b}})}]{Bretislav_1995}%
  \BibitemOpen
  \bibfield  {author} {\bibinfo {author} {\bibfnamefont {B.}~\bibnamefont
  {Friedrich}}\ and\ \bibinfo {author} {\bibfnamefont {D.}~\bibnamefont
  {Herschbach}},\ }\href {\doibase 10.1103/PhysRevLett.74.4623} {\bibfield
  {journal} {\bibinfo  {journal} {Phys. Rev. Lett.}\ }\textbf {\bibinfo
  {volume} {74}},\ \bibinfo {pages} {4623} (\bibinfo {year}
  {1995}{\natexlab{b}})}\BibitemShut {NoStop}%
\bibitem [{\citenamefont {Stapelfeldt}\ and\ \citenamefont
  {Seideman}(2003)}]{Stapelfeldt_2003}%
  \BibitemOpen
  \bibfield  {author} {\bibinfo {author} {\bibfnamefont {H.}~\bibnamefont
  {Stapelfeldt}}\ and\ \bibinfo {author} {\bibfnamefont {T.}~\bibnamefont
  {Seideman}},\ }\href {\doibase 10.1103/RevModPhys.75.543} {\bibfield
  {journal} {\bibinfo  {journal} {Rev. Mod. Phys.}\ }\textbf {\bibinfo {volume}
  {75}},\ \bibinfo {pages} {543} (\bibinfo {year} {2003})}\BibitemShut
  {NoStop}%
\bibitem [{\citenamefont {Torres}\ \emph {et~al.}(2005)\citenamefont {Torres},
  \citenamefont {de~Nalda},\ and\ \citenamefont {Marangos}}]{Torres_2005}%
  \BibitemOpen
  \bibfield  {author} {\bibinfo {author} {\bibfnamefont {R.}~\bibnamefont
  {Torres}}, \bibinfo {author} {\bibfnamefont {R.}~\bibnamefont {de~Nalda}}, \
  and\ \bibinfo {author} {\bibfnamefont {J.~P.}\ \bibnamefont {Marangos}},\
  }\href {\doibase 10.1103/PhysRevA.72.023420} {\bibfield  {journal} {\bibinfo
  {journal} {Phys. Rev. A}\ }\textbf {\bibinfo {volume} {72}},\ \bibinfo
  {pages} {023420} (\bibinfo {year} {2005})}\BibitemShut {NoStop}%
\bibitem [{\citenamefont {Ortigoso}\ \emph {et~al.}(1999)\citenamefont
  {Ortigoso}, \citenamefont {Rodr\'iques}, \citenamefont {Gupta},\ and\
  \citenamefont {B.Friedrich}}]{OrtigosoFriedrich99}%
  \BibitemOpen
  \bibfield  {author} {\bibinfo {author} {\bibfnamefont {J.}~\bibnamefont
  {Ortigoso}}, \bibinfo {author} {\bibfnamefont {M.}~\bibnamefont
  {Rodr\'iques}}, \bibinfo {author} {\bibfnamefont {M.}~\bibnamefont {Gupta}},
  \ and\ \bibinfo {author} {\bibnamefont {B.Friedrich}},\ }\href@noop {}
  {\bibfield  {journal} {\bibinfo  {journal} {Journal of Chemical Physics}\
  }\textbf {\bibinfo {volume} {110}},\ \bibinfo {pages} {3870} (\bibinfo {year}
  {1999})}\BibitemShut {NoStop}%
\bibitem [{\citenamefont {Lemeshko}\ \emph {et~al.}(2013)\citenamefont
  {Lemeshko}, \citenamefont {Krems}, \citenamefont {Doyle},\ and\ \citenamefont
  {Kais}}]{Lemeshko_2013_review}%
  \BibitemOpen
  \bibfield  {author} {\bibinfo {author} {\bibfnamefont {M.}~\bibnamefont
  {Lemeshko}}, \bibinfo {author} {\bibfnamefont {R.~V.}\ \bibnamefont {Krems}},
  \bibinfo {author} {\bibfnamefont {J.~M.}\ \bibnamefont {Doyle}}, \ and\
  \bibinfo {author} {\bibfnamefont {S.}~\bibnamefont {Kais}},\ }\href@noop {}
  {\bibfield  {journal} {\bibinfo  {journal} {Molecular Physics}\ }\textbf
  {\bibinfo {volume} {111}},\ \bibinfo {pages} {1648} (\bibinfo {year}
  {2013})}\BibitemShut {NoStop}%
\bibitem [{\citenamefont {Syvokon}(2006)}]{syvokon2006influence}%
  \BibitemOpen
  \bibfield  {author} {\bibinfo {author} {\bibfnamefont {V.}~\bibnamefont
  {Syvokon}},\ }\href@noop {} {\bibfield  {journal} {\bibinfo  {journal} {Low
  Temperature Physics}\ }\textbf {\bibinfo {volume} {32}},\ \bibinfo {pages}
  {48} (\bibinfo {year} {2006})}\BibitemShut {NoStop}%
\bibitem [{\citenamefont {Toennies}\ and\ \citenamefont
  {Vilesov}(2004)}]{ToenniesAngChem04}%
  \BibitemOpen
  \bibfield  {author} {\bibinfo {author} {\bibfnamefont {J.~P.}\ \bibnamefont
  {Toennies}}\ and\ \bibinfo {author} {\bibfnamefont {A.~F.}\ \bibnamefont
  {Vilesov}},\ }\href@noop {} {\bibfield  {journal} {\bibinfo  {journal}
  {Angewandte Chemie International Edition}\ }\textbf {\bibinfo {volume}
  {43}},\ \bibinfo {pages} {2622} (\bibinfo {year} {2004})}\BibitemShut
  {NoStop}%
\bibitem [{\citenamefont {Hoffman}\ \emph {et~al.}(2002)\citenamefont
  {Hoffman}, \citenamefont {McElroy}, \citenamefont {Lee}, \citenamefont
  {Lang}, \citenamefont {Eisaki}, \citenamefont {Uchida},\ and\ \citenamefont
  {Davis}}]{Hoffman}%
  \BibitemOpen
  \bibfield  {author} {\bibinfo {author} {\bibfnamefont {J.}~\bibnamefont
  {Hoffman}}, \bibinfo {author} {\bibfnamefont {K.}~\bibnamefont {McElroy}},
  \bibinfo {author} {\bibfnamefont {D.-H.}\ \bibnamefont {Lee}}, \bibinfo
  {author} {\bibfnamefont {K.}~\bibnamefont {Lang}}, \bibinfo {author}
  {\bibfnamefont {H.}~\bibnamefont {Eisaki}}, \bibinfo {author} {\bibfnamefont
  {S.}~\bibnamefont {Uchida}}, \ and\ \bibinfo {author} {\bibfnamefont
  {J.}~\bibnamefont {Davis}},\ }\href@noop {} {\bibfield  {journal} {\bibinfo
  {journal} {Science}\ }\textbf {\bibinfo {volume} {297}},\ \bibinfo {pages}
  {1148} (\bibinfo {year} {2002})}\BibitemShut {NoStop}%
\bibitem [{\citenamefont {Hanaguri}\ \emph {et~al.}(2007)\citenamefont
  {Hanaguri}, \citenamefont {Kohsaka}, \citenamefont {Davis}, \citenamefont
  {Lupien}, \citenamefont {Yamada}, \citenamefont {Azuma}, \citenamefont
  {Takano}, \citenamefont {Ohishi}, \citenamefont {Ono},\ and\ \citenamefont
  {Takagi}}]{hanaguri2007quasiparticle}%
  \BibitemOpen
  \bibfield  {author} {\bibinfo {author} {\bibfnamefont {T.}~\bibnamefont
  {Hanaguri}}, \bibinfo {author} {\bibfnamefont {Y.}~\bibnamefont {Kohsaka}},
  \bibinfo {author} {\bibfnamefont {J.}~\bibnamefont {Davis}}, \bibinfo
  {author} {\bibfnamefont {C.}~\bibnamefont {Lupien}}, \bibinfo {author}
  {\bibfnamefont {I.}~\bibnamefont {Yamada}}, \bibinfo {author} {\bibfnamefont
  {M.}~\bibnamefont {Azuma}}, \bibinfo {author} {\bibfnamefont
  {M.}~\bibnamefont {Takano}}, \bibinfo {author} {\bibfnamefont
  {K.}~\bibnamefont {Ohishi}}, \bibinfo {author} {\bibfnamefont
  {M.}~\bibnamefont {Ono}}, \ and\ \bibinfo {author} {\bibfnamefont
  {H.}~\bibnamefont {Takagi}},\ }\href@noop {} {\bibfield  {journal} {\bibinfo
  {journal} {Nature Physics}\ }\textbf {\bibinfo {volume} {3}},\ \bibinfo
  {pages} {865} (\bibinfo {year} {2007})}\BibitemShut {NoStop}%
\bibitem [{\citenamefont {Rutter}\ \emph {et~al.}(2007)\citenamefont {Rutter},
  \citenamefont {Crain}, \citenamefont {Guisinger}, \citenamefont {Li},
  \citenamefont {First},\ and\ \citenamefont
  {Stroscio}}]{rutter2007scattering}%
  \BibitemOpen
  \bibfield  {author} {\bibinfo {author} {\bibfnamefont {G.~M.}\ \bibnamefont
  {Rutter}}, \bibinfo {author} {\bibfnamefont {J.}~\bibnamefont {Crain}},
  \bibinfo {author} {\bibfnamefont {N.}~\bibnamefont {Guisinger}}, \bibinfo
  {author} {\bibfnamefont {T.}~\bibnamefont {Li}}, \bibinfo {author}
  {\bibfnamefont {P.}~\bibnamefont {First}}, \ and\ \bibinfo {author}
  {\bibfnamefont {J.}~\bibnamefont {Stroscio}},\ }\href@noop {} {\bibfield
  {journal} {\bibinfo  {journal} {Science}\ }\textbf {\bibinfo {volume}
  {317}},\ \bibinfo {pages} {219} (\bibinfo {year} {2007})}\BibitemShut
  {NoStop}%
\bibitem [{\citenamefont {Jeon}\ \emph {et~al.}(2014)\citenamefont {Jeon},
  \citenamefont {Zhou}, \citenamefont {Gyenis}, \citenamefont {Feldman},
  \citenamefont {Kimchi}, \citenamefont {Potter}, \citenamefont {Gibson},
  \citenamefont {Cava}, \citenamefont {Vishwanath},\ and\ \citenamefont
  {Yazdani}}]{jeon2014landau}%
  \BibitemOpen
  \bibfield  {author} {\bibinfo {author} {\bibfnamefont {S.}~\bibnamefont
  {Jeon}}, \bibinfo {author} {\bibfnamefont {B.~B.}\ \bibnamefont {Zhou}},
  \bibinfo {author} {\bibfnamefont {A.}~\bibnamefont {Gyenis}}, \bibinfo
  {author} {\bibfnamefont {B.~E.}\ \bibnamefont {Feldman}}, \bibinfo {author}
  {\bibfnamefont {I.}~\bibnamefont {Kimchi}}, \bibinfo {author} {\bibfnamefont
  {A.~C.}\ \bibnamefont {Potter}}, \bibinfo {author} {\bibfnamefont {Q.~D.}\
  \bibnamefont {Gibson}}, \bibinfo {author} {\bibfnamefont {R.~J.}\
  \bibnamefont {Cava}}, \bibinfo {author} {\bibfnamefont {A.}~\bibnamefont
  {Vishwanath}}, \ and\ \bibinfo {author} {\bibfnamefont {A.}~\bibnamefont
  {Yazdani}},\ }\href@noop {} {\bibfield  {journal} {\bibinfo  {journal}
  {Nature materials}\ }\textbf {\bibinfo {volume} {13}},\ \bibinfo {pages}
  {851} (\bibinfo {year} {2014})}\BibitemShut {NoStop}%
\bibitem [{\citenamefont {Lehnig}\ \emph {et~al.}(2009)\citenamefont {Lehnig},
  \citenamefont {Raston},\ and\ \citenamefont
  {J{\"a}ger}}]{lehnig_rotational_2009}%
  \BibitemOpen
  \bibfield  {author} {\bibinfo {author} {\bibfnamefont {R.}~\bibnamefont
  {Lehnig}}, \bibinfo {author} {\bibfnamefont {P.~L.}\ \bibnamefont {Raston}},
  \ and\ \bibinfo {author} {\bibfnamefont {W.}~\bibnamefont {J{\"a}ger}},\
  }\href {\doibase 10.1039/B819844F} {\bibfield  {journal} {\bibinfo  {journal}
  {Faraday Discuss.}\ }\textbf {\bibinfo {volume} {142}},\ \bibinfo {pages}
  {297} (\bibinfo {year} {2009})}\BibitemShut {NoStop}%
\bibitem [{\citenamefont {DeMarco}\ and\ \citenamefont
  {Jin}(1999)}]{deMarcoSci99}%
  \BibitemOpen
  \bibfield  {author} {\bibinfo {author} {\bibfnamefont {B.}~\bibnamefont
  {DeMarco}}\ and\ \bibinfo {author} {\bibfnamefont {D.~S.}\ \bibnamefont
  {Jin}},\ }\href@noop {} {\bibfield  {journal} {\bibinfo  {journal} {Science}\
  }\textbf {\bibinfo {volume} {285}},\ \bibinfo {pages} {1703} (\bibinfo {year}
  {1999})}\BibitemShut {NoStop}%
\bibitem [{\citenamefont {Leggett}(2006)}]{LeggettQuantLiquids}%
  \BibitemOpen
  \bibfield  {author} {\bibinfo {author} {\bibfnamefont {A.~J.}\ \bibnamefont
  {Leggett}},\ }\href@noop {} {\emph {\bibinfo {title} {Quantum Liquids: Bose
  Condensation and Cooper Pairing in Condensed-Matter Systems}}}\ (\bibinfo
  {publisher} {Oxford},\ \bibinfo {year} {2006})\BibitemShut {NoStop}%
\bibitem [{\citenamefont {G{\"u}nter}\ \emph {et~al.}(2006)\citenamefont
  {G{\"u}nter}, \citenamefont {St{\"o}ferle}, \citenamefont {Moritz},
  \citenamefont {K{\"o}hl},\ and\ \citenamefont {Esslinger}}]{GunterPRL06}%
  \BibitemOpen
  \bibfield  {author} {\bibinfo {author} {\bibfnamefont {K.}~\bibnamefont
  {G{\"u}nter}}, \bibinfo {author} {\bibfnamefont {T.}~\bibnamefont
  {St{\"o}ferle}}, \bibinfo {author} {\bibfnamefont {H.}~\bibnamefont
  {Moritz}}, \bibinfo {author} {\bibfnamefont {M.}~\bibnamefont {K{\"o}hl}}, \
  and\ \bibinfo {author} {\bibfnamefont {T.}~\bibnamefont {Esslinger}},\
  }\href@noop {} {\bibfield  {journal} {\bibinfo  {journal} {Phys. Rev. Lett.}\
  }\textbf {\bibinfo {volume} {96}},\ \bibinfo {pages} {180402} (\bibinfo
  {year} {2006})}\BibitemShut {NoStop}%
\bibitem [{\citenamefont {Levinsen}\ \emph {et~al.}(2015)\citenamefont
  {Levinsen}, \citenamefont {Parish},\ and\ \citenamefont
  {Bruun}}]{Levinsen_2015}%
  \BibitemOpen
  \bibfield  {author} {\bibinfo {author} {\bibfnamefont {J.}~\bibnamefont
  {Levinsen}}, \bibinfo {author} {\bibfnamefont {M.~M.}\ \bibnamefont
  {Parish}}, \ and\ \bibinfo {author} {\bibfnamefont {G.~M.}\ \bibnamefont
  {Bruun}},\ }\href {\doibase 10.1103/PhysRevLett.115.125302} {\bibfield
  {journal} {\bibinfo  {journal} {Phys. Rev. Lett.}\ }\textbf {\bibinfo
  {volume} {115}},\ \bibinfo {pages} {125302} (\bibinfo {year}
  {2015})}\BibitemShut {NoStop}%
\bibitem [{\citenamefont {Ngampruetikorn}\ \emph {et~al.}(2013)\citenamefont
  {Ngampruetikorn}, \citenamefont {Levinsen},\ and\ \citenamefont
  {Parish}}]{Ngampruetikorn_2013}%
  \BibitemOpen
  \bibfield  {author} {\bibinfo {author} {\bibfnamefont {V.}~\bibnamefont
  {Ngampruetikorn}}, \bibinfo {author} {\bibfnamefont {J.}~\bibnamefont
  {Levinsen}}, \ and\ \bibinfo {author} {\bibfnamefont {M.~M.}\ \bibnamefont
  {Parish}},\ }\href {\doibase 10.1103/PhysRevLett.111.265301} {\bibfield
  {journal} {\bibinfo  {journal} {Phys. Rev. Lett.}\ }\textbf {\bibinfo
  {volume} {111}},\ \bibinfo {pages} {265301} (\bibinfo {year}
  {2013})}\BibitemShut {NoStop}%
\bibitem [{\citenamefont {Parish}\ and\ \citenamefont
  {Levinsen}(2013)}]{Parish_2013}%
  \BibitemOpen
  \bibfield  {author} {\bibinfo {author} {\bibfnamefont {M.~M.}\ \bibnamefont
  {Parish}}\ and\ \bibinfo {author} {\bibfnamefont {J.}~\bibnamefont
  {Levinsen}},\ }\href {\doibase 10.1103/PhysRevA.87.033616} {\bibfield
  {journal} {\bibinfo  {journal} {Phys. Rev. A}\ }\textbf {\bibinfo {volume}
  {87}},\ \bibinfo {pages} {033616} (\bibinfo {year} {2013})}\BibitemShut
  {NoStop}%
\end{thebibliography}

\begin{thebibliography}{1}%
\makeatletter
\providecommand \@ifxundefined [1]{%
 \@ifx{#1\undefined}
}%
\providecommand \@ifnum [1]{%
 \ifnum #1\expandafter \@firstoftwo
 \else \expandafter \@secondoftwo
 \fi
}%
\providecommand \@ifx [1]{%
 \ifx #1\expandafter \@firstoftwo
 \else \expandafter \@secondoftwo
 \fi
}%
\providecommand \natexlab [1]{#1}%
\providecommand \enquote  [1]{``#1''}%
\providecommand \bibnamefont  [1]{#1}%
\providecommand \bibfnamefont [1]{#1}%
\providecommand \citenamefont [1]{#1}%
\providecommand \href@noop [0]{\@secondoftwo}%
\providecommand \href [0]{\begingroup \@sanitize@url \@href}%
\providecommand \@href[1]{\@@startlink{#1}\@@href}%
\providecommand \@@href[1]{\endgroup#1\@@endlink}%
\providecommand \@sanitize@url [0]{\catcode `\\12\catcode `\$12\catcode
  `\&12\catcode `\#12\catcode `\^12\catcode `\_12\catcode `\%12\relax}%
\providecommand \@@startlink[1]{}%
\providecommand \@@endlink[0]{}%
\providecommand \url  [0]{\begingroup\@sanitize@url \@url }%
\providecommand \@url [1]{\endgroup\@href {#1}{\urlprefix }}%
\providecommand \urlprefix  [0]{URL }%
\providecommand \Eprint [0]{\href }%
\providecommand \doibase [0]{http://dx.doi.org/}%
\providecommand \selectlanguage [0]{\@gobble}%
\providecommand \bibinfo  [0]{\@secondoftwo}%
\providecommand \bibfield  [0]{\@secondoftwo}%
\providecommand \translation [1]{[#1]}%
\providecommand \BibitemOpen [0]{}%
\providecommand \bibitemStop [0]{}%
\providecommand \bibitemNoStop [0]{.\EOS\space}%
\providecommand \EOS [0]{\spacefactor3000\relax}%
\providecommand \BibitemShut  [1]{\csname bibitem#1\endcsname}%
\let\auto@bib@innerbib\@empty
%</preamble>
\bibitem [{\citenamefont {Varshalovich}\ \emph {et~al.}(1988)\citenamefont
  {Varshalovich}, \citenamefont {Moskalev},\ and\ \citenamefont
  {Khersonskii}}]{Varshalovich1}%
  \BibitemOpen
  \bibfield  {author} {\bibinfo {author} {\bibfnamefont {D.~A.}\ \bibnamefont
  {Varshalovich}}, \bibinfo {author} {\bibfnamefont {A.}~\bibnamefont
  {Moskalev}}, \ and\ \bibinfo {author} {\bibfnamefont {V.}~\bibnamefont
  {Khersonskii}},\ }\href@noop {} {\emph {\bibinfo {title} {Quantum theory of
  angular momentum}}}\ (\bibinfo  {publisher} {World Scientific},\ \bibinfo
  {year} {1988})\BibitemShut {NoStop}%
\end{thebibliography}

%

\newpage

\end{widetext}

\end{document}